\documentclass[a4paper]{article}

\usepackage{microtype}

\usepackage{graphicx}
\usepackage{epstopdf}
\usepackage[numbers]{natbib}

\usepackage{amsmath}
\usepackage{amssymb}

\usepackage{ifthen}
\newboolean{extendedabstract}

\setboolean{extendedabstract}{false}

\newcommand{\abstractonly}[1]{
\ifthenelse{\boolean{extendedabstract}}{#1}{}
}

\newcommand{\fullonly}[1]{
\ifthenelse{\boolean{extendedabstract}}{}{#1}
}

\newcommand{\NN}{\mathbb{N}}

\abstractonly{\title{A New Perspective on the Windows Scheduling Problem (Extended Abstract\footnote{ Proofs of lemmas and theorems are omitted or sketched in this extended abstract. The full version of this work is appended.})}}
\fullonly{\title{A New Perspective on the \\Windows Scheduling Problem 
}}

\usepackage{authblk}

\author{Tobias Jacobs}
\author{Salvatore Longo}
\affil{
NEC Laboratories Europe\\
Kurf\"ursten-Anlage 36\\
69115 Heidelberg, Germany\\
\texttt{\{tobias.jacobs,salvatore.longo\}@neclab.eu}
}

\newtheorem{theorem}{Theorem}
\newtheorem{lemma}[theorem]{Lemma}
\newtheorem{corollary}[theorem]{Corollary}

\newtheorem{example}[theorem]{Example}
\newtheorem{definition}[theorem]{Definition}

\begin{document}


\date{October 2014}
\maketitle

\begin{abstract}
The Windows Scheduling Problem, also known as the Pinwheel Problem, is to schedule periodic jobs subject to their processing frequency demands. Instances are given as a set of jobs that have to be processed infinitely often such that the time interval between two consecutive executions of the same job $j$ is no longer than the job's given period~$p_j$.

The key contribution of this work is a new interpretation of the problem variant with \emph{exact periods}, where the time interval between consecutive executions must be strictly $p_j$. We show that this version is equivalent to a natural combinatorial problem we call \emph{Partial Coding}. Reductions in both directions can be realized in polynomial time, so that both hardness proofs and algorithms for Partial Coding transfer to Windows Scheduling.

Applying this new perspective, we obtain a number of new results regarding the computational complexity of various Windows Scheduling Problem variants. We prove that even the case of one processor and unit-length jobs does not admit a pseudo-polynomial time algorithm unless SAT can be solved by a randomized method in expected quasi-polynomial time. This result also extends to the case of inexact periods, which answers a question that has remained open for more than two decades. Furthermore, we report an error found in a hardness proof previously given for the multi-machine case without machine migration, and we show that this variant reduces to the single-machine case.
Finally, we prove that even with unit-length jobs the problem is co-NP-hard when jobs are allowed to migrate between machines.

\end{abstract}



\section{Introduction}
\label{sec:intro}

We address the problem of scheduling periodic jobs subject to processing frequency demands, which is commonly referred to as \emph{Windows Scheduling} or \emph{Pinwheel Scheduling}. 
Given a set of jobs $j = 1,\ldots,n$ with \emph{periods} $p_1,\ldots,p_n$, the objective is to decide whether there exists an infinite non-preemptive schedule where for each job $j$ the time between any two subsequent executions of $j$ is no longer than $p_j$. Equivalently, in any time interval of length $p_j$ an execution of job $j$ must be started at least once.

Windows Scheduling has a number of applications in various domains. Consider for example a service team that needs to do maintenance work on a number of objects (like rooms, buildings, or technical appliances), and for each of the objects a service contract specifies the minimum frequency of maintenance. Here Windows Scheduling appears as the problem to compute the team's work plan, and to decide whether a new service contract can be handled by the team on top of the existing contracts.

In another important class of applications, regular messages from multiple senders have to be transmitted through a time-multiplexed communication channel, where each sender needs to transmit its messages with a specific frequency. Windows Scheduling appears here as the task to assign transmission time intervals to senders. A popular scenario that has motivated studies of the problem is where multiple satellites send information updates to a ground station~\cite{holte89}. In other scenarios, frames of multiple videos need to be sent to clients of a video-on-demand service, or readings from multiple sensors need to be transmitted in regular intervals to a data storage.

\noindent
\textbf{Problem variants.} We give a short taxonomy of the problem variants that are most relevant for this work.
We speak of \emph{unit-length} jobs when each job has a processing time of 1; otherwise the length $\ell_j \in \NN$ of each job $j$ is given as part of the input. Furthermore, we distinguish between the single-machine case and two multi-machine variants. In the first multi-machine variant, job migration is disallowed, which means that the schedule needs to assign each  job to a fixed machine where all executions of the job then take place. In the second variant, job migration is allowed, so each execution of a job can take place on a different machine. Finally, we distinguish between exact periods and inexact periods. In case of exact periods, the time interval between any two subsequent executions of the same job $j$ has to have \emph{exactly} length $p_j$, whereas in the inexact case $p_j$ is an upper bound on the length of that time interval. 

An important parameter characterizing instances of the Windows Scheduling Problem is the \emph{density} of the instance. Given job periods $p_1,\ldots,p_n$ and job lengths $\ell_1,\ldots,\ell_n$, the density of the problem instance is defined as $\sum_j \frac{\ell_j}{p_j}$. Intuitively, the share of all available processing time required by a job $j$ is $\frac{\ell_j}{p_j}$, and the density is calculated as the sum over these fractions. A known necessary condition for problem instances to admit a feasible solution is that the density must be no more than~$1$ (see~\cite{holte89}).

\noindent
\textbf{Related work.} The class of problems we are addressing has been referred to in literature as Pinwheel Scheduling, Windows Scheduling, or Periodic Scheduling. The term Pinwheel Scheduling has been coined by Holte et al. in~\cite{holte89} to describe the problem to schedule $n$ unit-length jobs on a single machine subject to inexact period requirements. The authors have shown that Pinwheel instances having a density of $0.5$ or less always admit a feasible schedule that can be computed in polynomial time. This density bound has been improved to $\frac{2}{3}$ and then $0.7$ by Chan and Chin~\cite{chan93,chan92}.

Algorithms for some special cases of the Pinwheel were also presented in~\cite{holte89}. Among these special cases are \emph{harmonic} instances where $p_i$ is a multiple of $p_j$ for any two jobs $i,j$ with $p_i<p_j$, and  instances having of up to $3$ distinct period lengths. For the case of $2$ distinct periods, an algorithm that finds the cyclic schedule with the minimum cycle length has been given by Holte et al. in~\cite{holte92}.

Korst et al. have addressed the Windows Scheduling Problem with exact periods and non-uniform job lengths~\cite{korst91}. They have shown that the multi-machine problem is strongly NP-hard, regardless of whether machine migration is allowed or not. They have further proven that even for a given schedule the problem of assigning job executions to machines is NP-hard, again both for the case with machine migration and the case where the assignment of jobs to machines has to be fixed. The authors also propose a number of approximation algorithms for the problem to minimize the number of machines. In~\cite{Eisenbrand10} Eisenbrand et al. give a factor 2 approximation for harmonic instances and show that this is the best achievable under standard complexity-theoretic assumptions. It is further shown in their article that the general minimization problem does not admit a factor $n^{1 - \epsilon}$ approximation for any $\epsilon>0$, where $n$ is the number of jobs.

The Windows Scheduling Problem with exact periods has also been studied under the name \emph{Perfectly Periodic Schedules}, which typically refers to the  problem to minimize the deviation of the schedule from the demanded job periods. This optimization problem has been addressed in~\cite{bar02,bar02a,bar04,brakerski03,brakerski02,patil06}. 

The single-machine version the of Windows Scheduling Problem with exact periods and non-uniform job lengths has been addressed by Korst et al. in~\cite{korst96}. Here it is shown that even the single-machine case is strongly NP-hard. Furthermore, the authors have generalized a result from \cite{holte89} proving that the special case where both the periods and the execution times are pairwise divisible admits a polynomial time algorithm.

What is referred to as Generalized Pinwheel Scheduling in literature is Windows Scheduling with inexact periods, a single machine, and non-uniform job lengths.  This version has been studied by Feinberg et al. in~\cite{feinberg05}. In that work, necessary conditions for feasibility based on problem relaxations were given, and an experimental study was conducted to evaluate several heuristics. The multi-machine case of that version has been addressed by Bar-Noy et al. in~\cite{bar03}. The authors give an algorithm that finds the asymptotically minimum number of machines to schedule a given set of jobs and also study the problem to maximize the number of jobs that can be processed on a given number of machines. In~\cite{bar07} the problem is interpreted as a restricted version of Bin Packing. Approximations for the task to minimize the number of machines for online and offline versions have been given in~\cite{bar07,bar12}. 

There has been some degree of confusion about the computational complexity of Windows Scheduling with unit-length jobs. In~\cite{holte89} it was claimed that a compact representation of the problem with inexact periods is NP-hard, but, as observed by the authors of~\cite{feinberg05}, the claim was not proven in that work. A hardness proof for the single machine case with exact periods was given by Bar-Noy et al. in~\cite{bar02}, and another hardness proof for the more general multi-machine version was given by Bar-Noy et al. in~\cite{bar07}. The latter proof is obsolete in light of the hardness proof for the less general problem in~\cite{bar02}, and the reduction additionally contains an error as we point out in Section~\ref{sec:hardness}. In~\cite{bar07}  the authors also show that the hardness of the exact periods case implies hardness of a compact representation of the version with inexact periods. This implication holds as well for the single-machine case. To summarize, a correct proof of NP-hardness for the compact representation can be assembled from results available in literature, although we are not aware of any work where this has been pointed out and belayed with the correct citations.

\noindent
\textbf{Our contributions.} We look at the Windows Scheduling Problem from a new perspective by interpreting it as \emph{Partial Coding}. The latter is a natural and intuitive combinatorial problem which we believe is of independent interest. Roughly speaking, the objective of Partial Coding is to assign coding vectors to objects such that each object can be unambiguously identified, despite the fact that the assigned object codes can only partially be read correctly.
This interpretation, which is introduced in Section~\ref{sec:pc}, reduces the Windows Scheduling Problem with exact periods to its \emph{combinatorial core}, i.e., number-theoretic concepts like modulo operators or greatest common divisors are no longer needed in the~analysis.

From this perspective we derive a number of new results regarding the computational tractability of various Windows Scheduling variants.
It has been shown prior to this work that the case of unit-length jobs with exact periods and a single machine is weakly NP-hard. We tighten this result by showing that this case does not even admit a pseudo-polynomial time algorithm unless SAT can be solved by a randomized method in expected time $n^{O(\log n \log \log n)}$. We further show that the same complexity also holds for inexact periods, which answers to a large extent the long-standing open question on whether the Pinwheel problem in standard representation admits an efficient algorithm~\cite{bar07,holte89}. These results are presented in Section~\ref{sec:hardness}.

In Section~\ref{sec:multimachine} we present our findings regarding the multi-machine case with exact periods.  We first identify an error in a previously given~\cite{bar07} NP-hardness proof for multi-machine Windows Scheduling with unit length jobs without machine migration. We then show that  the multi-machine Windows Scheduling problem without machine migration can be reduced to the single-machine case. The section is concluded with a proof that allowing machine migration makes the problem co-NP-hard, which implies that this particular problem version is neither in NP nor in co-NP unless these two complexity classes are equal.

\section{Partial Coding}
\label{sec:pc}

In this section we introduce the \emph{Partial Coding (PC) Problem} and establish its relationship with Windows Scheduling.

In instances of Partial Coding we are given a set of $d$ \emph{attributes}. Each attribute $i = 1,\ldots,d$ can assume integer values from the interval $[0,f_i-1]$, where $f_i \geq 2$ is given as the \emph{value range} of the $i$th attribute. Values of these attributes will be assigned in order  to encode a given family of $n$ \emph{symbols} $s_1,\ldots,s_n$. Each symbol is associated with a subset of attributes available for its encoding, that is, $s_j \subseteq \{1,\ldots,d\}$ for $j=1,\ldots,n$.

Formally, an encoding of a symbol $s_j$ is a mapping $c_j: s_j \to \NN$, such that $c_j(i) \in [0,f_i-1]$, so the encoding assigns an integer value to each attribute that is contained by the symbol $s_j$. A collection $c_1,\ldots,c_n$ of encodings of the $n$ symbols is \emph{feasible} if any pair of symbols can be distinguished based on the values of the attributes the two symbols have in common. More specifically, for the encoding to be feasible, there needs to be some attribute $i \in s_j \cap s_k$ with $c_j(i) \neq c_k(i)$ for any index pair $j \neq k$.
\begin{definition}
Given $d$ attributes with value ranges $f_1,\ldots,f_d$ and a family of $n$ symbols $s_1,\ldots,s_n \subseteq \{1,\ldots,d\}$, the Partial Coding (PC) Problem is to decide whether there is a feasible encoding of the symbols in terms of the attributes.
\end{definition}
We refer to attributes with value range $\{0,1\}$ as \emph{binary attributes}. In case that all attributes are binary we have an instance of the \emph{Binary Partial Coding (BPC)} Problem.

\begin{figure}[h]
\begin{minipage}{0.5 \textwidth}
\center
\begin{tabular}{l|cccc}
{} & $s_1$ & $s_2$ & $s_3$ & $s_4$
\\
\hline
\{0,1\} & $$ & $\times$ & $$ & $$
\\
\{0,1,2\} & $\times$ & $\times$ & $$ & $$
\\
\{0,1\} & $$ & $$ & $$ & $$
\end{tabular}
\end{minipage}
\begin{minipage}{0.5 \textwidth}
\center
\begin{tabular}{l|cccc}
{} & $s_1$ & $s_2$ & $s_3$ & $s_4$
\\
\hline
{\{0,1\}} & $0$ & $\times$ & $1$ & $1$
\\
{\{0,1,2\}} & $\times$ & $\times$ & $0$ & $2$
\\
{\{0,1\}} & $0$ & $1$ & $0$ & $0$
\end{tabular}
\end{minipage}
\caption{
\label{fig:pcexample}
An instance of the Partial Coding Problem with 4 symbols and 3 binary attributes is shown on the left-hand side; a feasible solution to it is given on the right-hand side. The problem instance would be formally written as $d=3$, $f_1 = f_3 = 2$, $f_2 = 3$, $n=4$, $s_1 = \{1,3\}$, $s_2 = \{3\}$, $s_3 = s_4 = \{1,2,3\}.$
}
\end{figure}

\begin{example}
An example instance with $d=3$ binary attributes and $n=4$ symbols is given in Figure~\ref{fig:pcexample}, which also shows a convenient way to graphically represent Partial Coding instances. On the left-hand side of the figure the problem instance is represented as a $d \times n$ matrix with columns representing symbols and rows representing attributes. Matrix cells are crossed out whenever the symbol corresponding to the column does not contain the attribute corresponding to the row. The objective is to populate the non-crossed matrix entries with integers such that each row contains only integers within the row's range, and for each pair of columns there exists some row where the two entries are not crossed and differ from each other. A feasible solution to the instance is shown on the right-hand side of the figure.
\end{example}

Partial Coding has applications independent from Windows Scheduling. Assume for example electronic identifier tags that are attached to objects. Each tag can be configured to carry a number consisting of $d$ digits. Some of the $d$ memory cells of some tags are known to be faulty, that is, the digit they return when reading them is not always the digit previously written. The problem to configure the tags such that each object can be unambiguously be identified in spite of faulty memory cells is an application of Partial Coding.

In the remainder of this section we establish the connection between the Windows Scheduling Problem and the Partial Coding Problem. This will be achieved using a class of factorizations of the set of job periods. We start by introducing the underlying number-theoretic concepts.

\begin{definition}
A vector of strictly positive integers $(f_1,\ldots,f_d)$ with $f_1 \leq \ldots 
\leq f_d$ is called a relative prime vector if for any $1 \leq i,j \leq d$ either $f_i = f_j$ or $f_i$ and $f_j$ are relatively prime.
\end{definition}

\begin{example} The vector $(5,7,12)$ is a relative prime vector, but $(5,9,12)$ is not a relative prime vector because $9$ and $12$ have $3$ as a common prime factor. However, $(5,5,12)$ again is a relative prime vector as $12$ and $5$ are relatively prime and the first two vector components are equal.
\end{example}

Given a positive integer $m$ and an integer exponent $h$, the residue $x \bmod m^h$ of any integer $x$ can be represented as a vector of $h$ integers between $0$ and $m-1$, defined as $( v_0(x), \ldots, v_{h-1}(x) )$ with
$$
v_i(x) = \left\lfloor \frac{x}{m^i} \right\rfloor \bmod m
$$
for $i = 0,\ldots,h-1$. We call $(v_0(x),\ldots,v_{h-1}(x))$ the \emph{base $m$ representation of $x \bmod m^h$}. 

The following lemma \abstractonly{(see full paper for a proof)}
states that $x \bmod m^h$ can indeed be reconstructed from its base $m$ representation.

\begin{lemma}
\label{lemma:basem}
For any integers $x$ and $m,h \geq 1$ it holds that 
$$
x \bmod m^h = \sum_{i=0}^{h-1} v_i(x) m^i = \sum_{i=0}^{h-1} \left( \left\lfloor \frac{x}{m^i} \right\rfloor \bmod m \right) m^i 
$$
\end{lemma}
\fullonly{
\textbf{Proof.} From the base $m$ representation of $x$, 
$$
x = \sum_{i=0}^\infty \left( \left\lfloor \frac{x}{m^i} \right\rfloor \bmod m \right) m^i \ ,
$$
follows that
$$
x \bmod m^h = \left( \sum_{i=0}^{h-1} \left( \left\lfloor \frac{x}{m^i} \right\rfloor \bmod m  \right) m^i+  \sum_{i=h}^\infty \left( \left\lfloor \frac{x}{m^i} \right\rfloor \bmod m   \right) m^i \right) \bmod m^h \ .
$$
Each summand of the right-hand sum is a multiple of $m^h$ and therefore $0$ modulo $m^h$. The left-hand sum adds up to a number less than $m^h$ and therefore it is not altered by the outer modulo operator, implying the claim.
\hfill{$\Box{}$}
}
\begin{example}
Let $m=3$ and $h=4$. The base $3$ representation of $345 \bmod 3^4 = 21$ is 
 $(345 \bmod 3, \lfloor 345 / 3 \rfloor \bmod 3, \lfloor 345 / 9 \rfloor \bmod 3, \lfloor 345 / 27 \rfloor \bmod 3) = (0,1,2,0)$.
Along the lines of Lemma~\ref{lemma:basem}, the value of $245 \bmod 3^4$ can be recomputed as $0 \cdot 3^0 + 1 \cdot 3^1 + 2 \cdot 3^2 + 0 \cdot 3^3 = 21$.
\end{example}

We next define the residue vector of an integer as the combination of base $m$ representations for several values of $m$.

\begin{definition}
\label{def:residuevector}
Given a relative prime vector $F = (f_1,\ldots,f_d)$ and an integer $x$, the residue vector of $x$ with respect to $F$ is defined as $R(x) := (r_1(x),\ldots,r_d(x)) $ with
$$
r_i (x) = \left\lfloor \frac{x}{f_i^h} \right\rfloor \bmod f_i  \ \text{with} \ h:= \max\{i-j \mid f_j = f_i \} \ .
$$
\end{definition}
It follows that for any factor $f$ appearing in the relative prime vector $h$ times, the residue vector $R(x)$ contains a base $f$ representation of $x \bmod f^h$ as a contiguous sub-vector.

\begin{example}
Consider the relative prime vector $f = (5,5,6)$. According to Definition~\ref{def:residuevector} the residue vector of $x = 345$ with respect to $f$ is $(x \bmod 5, \lfloor 354 / 5 \rfloor \bmod $5$, x \bmod 6) = (0,4,3)$. Observe that the partial residue vector $(3)$ is the base $3$ representation of $x \bmod 3 = 2$, and the partial vector $(0,4)$ is the base $5$ representation of $x \bmod 5^2 = 20$. 
\end{example}

 We next give a generalization of Lemma~\ref{lemma:basem}, showing that this combination of representations is a unique representation of the residue with respect to the product of members of~$F$.

\begin{lemma}
\label{lemma:residuevector}
Given a relative prime vector $F = (f_1,\ldots,f_d)$ with $\prod_i f_i = n$, the residue class of $x$ modulo $n$ is uniquely characterized by the residue vector $R(x)$ with respect to $F$, that is, for any integers $x,y$ it holds that $R(x) = R(y)$ if and only if it holds that $x \equiv y \ (\bmod n)$.
\end{lemma}
\fullonly{
\textbf{Proof.} For the \emph{if} part, observe that, for $i = 1,\ldots,d$, 
$$
	r_i(x \bmod n) =  \left\lfloor \frac{x \bmod n}{f_i^h} \right\rfloor \bmod f_i = \left\lfloor \frac{x}{f_i^h} \right\rfloor \bmod n \bmod f_i = r(x) \ ,
$$
where the second equation follows from $n$ being a multiple of $f_i^h$, and the third equation follows from $n$ being multiple of $f_i$. Consequently, $x \equiv y\ (\bmod n\ )$ implies that $R(x) = R(x \bmod n) = R(y \bmod n) = R(y)$.

For the \emph{only if} part, we first consider the special case where the members of $F$ are pairwise distinct, that is, $f_i \neq f_j$ for any $i \neq j$. In that case the residue vector $R(x)$ reduces to $R(x) = (x \bmod f_1, \ldots, x \bmod f_d)$. For any 
two integers $x$ and $y$ with $R(x) = R(y)$, the difference $x-y$ must be a multiple of $f_i$ for each $i = 1,\ldots,d$. As $F$ is a relative prime vector with pairwise distinct elements, it follows that $x-y$ is a multiple of $\prod_i f_i = n$ and therefore $x \equiv y\ (\bmod n)$.

For the general case, assume that $F$ contains $e \leq d$ distinct factors $g_1,\ldots,g_e$, and assume that $g_i$ appears $h_i$ times in $F$ for $i=1,\ldots,e$. 
Consider any specific factor $g_i$ and let $f_j = \ldots = f_k$ be the $h_i$ components of $F$ equal to $g_i$. As these $h_i$ factors constitute the base $g_i$ representation of $g_i^{h_i}$, we can apply  Lemma~\ref{lemma:basem} to show that that $R(x) = R(y)$ implies $x \equiv y\  (\bmod {g_i}^{h_i})$ for each $i = 1,\ldots,e$. Clearly, $\prod_{i=1}^e g_i^{h_i} = n$, and $g_1^{h_1},\ldots,g_e^{h_e}$ are pairwise prime, so the same reasoning as in the preceding special case can be applied to show that $R(x) = R(y)$ implies  $x \equiv y\ (\bmod n)$.
\hfill{$\Box{}$}
}
\abstractonly{
\textbf{Proof sketch.} For the \emph{if} part one can show that $r_i(x \bmod n) = r_i(x)$.  Consequently, $x \equiv y\ (\bmod n\ )$ implies that $R(x) = R(x \bmod n) = R(y \bmod n) = R(y)$.
For the \emph{only if} part, we first consider the special case where the members of $F$ are pairwise distinct, so that the residue vector $R(x)$ reduces to $R(x) = (x \bmod f_1, \ldots, x \bmod f_d)$. In that case $R(x) = R(y)$ implies that $x-y$ is a multiple of each $f_i$ and therefore also a multiple of  $\prod_i f_i = n$. In the general case where any factor $g$ can appear $h$ times in $F$ for some $h \geq 1$, we use the fact that $R(x)$ contains a base $g$ representation of $g^h$, so Lemma~\ref{lemma:basem} can be applied to show that $R(x) = R(y)$ implies $x \equiv y\  (\bmod g^h)$. This reduces the general case to the previous special case.
\hfill{$\Box{}$}
}

We have shown how to represent residue classes in terms of residue vectors with respect to relative prime vectors. The final step to make this concept applicable to the Windows Scheduling Problem is the following definition.

\begin{definition}
\label{def:rpf}
Given a multi-set of strictly positive integers $P = \{p_1,\ldots,p_n\}$, a relative prime factorization of $P$ is a relative prime vector  $F = (f_1,\ldots,f_d)$ where for each $p_j \in P$ there is some subset $s_j \subseteq \{1,\ldots,d\}$ with $\prod_{i \in s_j} f_i = p_j$.
\end{definition}

\begin{example}
Consider the set $P = (30,42)$. The prime factorizations of these numbers give one possible relative prime factorization $(2,3,5,7)$. An alternative relative prime factorization is $(5,6,7)$.
\end{example}

Like in the above example, one particular relative prime factorization of $P$ can always be obtained as the union of all prime factorizations of the members of $P$. However, such factorizations are computationally problematic as no efficient algorithm for prime factorization is known. Other relative prime factorizations can be computed efficiently as we show below.

We are now ready to use the number-theoretic concepts just introduced to show the equivalence of Windows Scheduling and Partial Coding. 

\begin{theorem}
\label{thm:wstopc}
The Windows Scheduling Problem can be reduced to the Partial Coding Problem in polynomial time.
\end{theorem}

Consider an instance of the Windows Scheduling Problem with the set of periods $P = \{p_1,\ldots,p_n\}$. The algorithm used for the reduction will compute a relative prime factorization  $F = (f_1,\ldots,f_d)$ of the set of periods. This relative prime factorization will serve as the set of attributes of the Partial Coding Problem instance. For each job $j$ in the scheduling instance, there will be one symbol to be encoded, and the attributes of that symbol will correspond to a set of factors of $F$ that appear in $p_j$. The first step of the reduction therefore is to give a polynomial time algorithm to compute the relative prime factorization. 

\begin{lemma}
Relative prime factorizations can be computed in polynomial time.
\end{lemma}
\fullonly{
\textbf{Proof.} The algorithm starts by assigning $F := P$ and applies in each iteration  the same  operation to some suitable pair of elements of $F$. The algorithm uses the operator $\textrm{split}:\NN^2 \to \NN^3$, defined as 
$$
\textrm{split}(x,y) := \left( \frac{x}{\textrm{gcd}(x,y)}, \textrm{gcd}(x,y), \frac{y}{\textrm{gcd}(x,y)} \right) \ ,
$$
where $\text{gcd}(x,y)$ is the greatest common divisor of $x$ and $y$.

In each iteration, the operator is applied to some pair of numbers $x,y \in F$ with $\textrm{gcd}(x,y) \neq 1$ and $x \neq y$. The two numbers $x,y$ are then replaced in $F$ with the three numbers computed by the split operator. Every integer that could be represented by a product of members of $F$ before some iteration can also be represented afterwards, so the invariant that every member of $P$ can be represented as such a product is preserved throughout the execution of the algorithm. The algorithm terminates as soon as there is no pair $x,y$ left to which the split operator can be applied. After sorting the elements of $F$ we arrive at a relative prime factorization of $P$. 

To upper bound the runtime of the algorithm, observe that in each iteration the product of all elements of $F$ decreases by factor $\text{gcd}(x,y) \geq 2$, and therefore the number of iterations is upper bounded by $\sum_{i=1}^n \log p_i$, which is polynomial in the representation size of the scheduling instance. As each individual iteration can be realized in polynomial time, the overall runtime of the algorithm is polynomial.
\hfill{$\Box{}$}
}
\abstractonly{
\textbf{Proof sketch.} The algorithm starts by assigning $F := P$ and applies in each iteration  the same  operation to some suitable pair of elements of $F$. The algorithm uses the operator $\textrm{split}:\NN^2 \to \NN^3$, defined as 
$$
\textrm{split}(x,y) := \left( \frac{x}{\textrm{gcd}(x,y)}, \textrm{gcd}(x,y), \frac{y}{\textrm{gcd}(x,y)} \right) \ ,
$$
where $\text{gcd}(x,y)$ is the greatest common divisor of $x$ and $y$.

In each iteration the operator is applied to some pair of numbers $x,y \in F$ with $\textrm{gcd}(x,y) \neq 1$ and $x \neq y$. The two numbers $x,y$ are then replaced in $F$ with the three numbers computed by the split operator. The number of iterations needed to arrive at a relative prime factorization is bounded by the number of prime factors contained by the members of~$P$.
\hfill{$\Box{}$}
}
\begin{example}
\sloppy
In case of $P = (30,42)$ the algorithm needs only one iteration where it replaces the pair $(30,42)$ with the triplet $(30 / \textrm{gcd}(30,42),\textrm{gcd}(30,42), 42 / \textrm{gcd}(30,42)) = (5,6,7)$, arriving at a relative prime factorization.
\end{example}

\fussy
We now go into details about how to determine the $n$ symbols of the Partial Coding Problem instance from a given relative prime factorization $F = (f_1,\ldots,f_d)$ of the set of periods. For each job $j$ with period $p_j$, we assign $s_j \subseteq \{1,\ldots,d\}$ to be the lexicographically smallest set satisfying $\prod_{i \in s_j} f_i = p_j$. As the components of $F$ are nondecreasingly sorted and pairwise either equal or prime, the requirement to be the lexicographically smallest set only has an effect on the choice between non-distinct elements of $F$. For example, if $F = (4,5,5)$ and $p_j = 20$, then both $f_1 f_2 = p_i$ and $f_1 f_3 = p_i$, but $s_j = \{1,2\}$ due to the requirement to be lexicographically smallest.

The proof that the two problem instances are equivalent will make use of a necessary and sufficient condition for pairs of jobs to collide in a schedule that is well-known in the context of Windows Scheduling (see e.g. \cite{bar02,wei83}). 
\fullonly{We give a proof below for the sake of completeness.}
A collision of two jobs is the situation when they have to be executed at the same time. 

\begin{lemma}
\label{lemma:collision}
Let $i,j$ be two distinct jobs having unit length and strict periods $p_i$ and $p_j$, and let $t_i$ and $t_j$ be a start time of $i$ and $j$ in a one-machine schedule, respectively. The two jobs do not collide if and only if $t_i \not \equiv t_j \pmod {\gcd\{p_i,p_j\}}$.
\end{lemma}
\fullonly{
\textbf{Proof.} Let $g := gcd(p_i,p_j)$. If the two jobs collide, then there is some time $t^*$ with $t_i + k_i p_i = t^* = t_j + k_j p_j$ for some integers $k_i, k_j$. As both $p_i$ and $p_j$ are multiples of $g$, it follows that
$$
t_i \bmod g = t^* \bmod g = t_j \bmod g \ .
$$
For showing the reverse implication, assume that $t_i \bmod g = t_j \bmod g =: y$ and let $r_i := p_i / g$ and $r_j := p_j / g$. Let further $z_i := \lfloor t_i / g \rfloor$ and $z_j := \lfloor t_j / g \rfloor$. Due to the definition of the greatest common divisor, $r_i$ and $r_j$ are relatively prime and therefore there are integers $b_i,b_j$ such that $z_i + b_i r_i = z_j + b_j r_j$. It follows that
\begin{align*}
{} & (z_i + b_i r_i) g + y = (z_j + b_j r_j) g + y  \\
\Rightarrow \ & (g z_i  + y) + b_i  r_i g  = (g z_j  + y) + b_j r_j  g\\
\Rightarrow \ & t_i + b_i  p_i  = t_j + b_j p_j \ ,
\end{align*}
which means that the jobs $i$ and $j$ collide at time $t_i + b_i  p_i$.
\hfill{$\Box{}$}
}

\fullonly{
\noindent
\textbf{Proof of Theorem~\ref{thm:wstopc}}. Using the above method to compute a Partial Coding Problem instance with value ranges $(f_1,\ldots,f_d)$ and symbols $s_1,\ldots,s_n$ from a Windows Scheduling Problem instance with periods $p_1,\ldots,p_n$, it remains to show that there is a feasible solution to the scheduling problem if and only if there is a feasible partial coding.

Due to the exact periods, any schedule is completely determined by a start time $t_j \in [0,p_j-1]$ for each job $j=1,\ldots,n$. For any schedule, whether feasible or not, we define the corresponding partial coding of the symbols as the residue vectors with respect to $F$ of the start times. Formally, for $j=1,\ldots,n$ symbol $s_j$ is assigned to encoding $c_j$ defined as 
$$
c_j(i) = r_i(t_j) \ ,
$$
where $r_i(t_j)$ is the $i$th component of the residue vector $R(t_j)$ with respect to $F$. 

For each $j=1,\ldots,n$ let $R_j(t_j)$ be the partial vector of $R(t_j)$ that is composed of all components $r_i(t_j)$ with $i \in s_j$. This vector contains exactly all the attribute values of the encoding $c_j$ of symbol $s_j$. Furthermore, $R_j(t_j)$ is the residue vector of $t_j$ with respect to $F_j$, which is defined as the relative prime vector containing all components of $F$ whose indices appear in $s_j$. By definition of $s_j$, the product of all components of $F_j$ is $p_j$. Using  Lemma~\ref{lemma:residuevector}, it follows that $t_j$ is uniquely determined by the attribute values of the encoding of symbol $s_j$. As the number of different attribute value combinations for symbol $s_j$ is $\prod_{i \in s_j} f_i = p_j$, our function mapping start times $t_j$ to encodings $c_j$ is is a bijection.

The theorem is proved by showing that the encoding given as $c_1,\dots,c_n$ is feasible if and only if the schedule given as $t_1,\ldots,t_n$ is feasible. As the schedule is feasible if and only if there is no pair of colliding jobs, and the encoding is defined to be feasible if any pair of symbols can be distinguished, it suffices to show that two jobs $j,k$ collide if and only if the symbols $s_j,s_k$ cannot be distinguished. Using Lemma~\ref{lemma:collision} and the way the encoding is determined from $t_j,t_k$, we only need to show that $t_j \equiv t_k\ (\bmod\ \text{gcd}(p_j,p_k))$ if and only if $ r_i(t_j) = r_i(t_k)$ for each $i \in s_j \cap s_k$. 

Let $s_{jk} := s_j \cap s_k$, and observe that 
\begin{equation}
\label{eq:gcd}
\text{gcd}(p_j,p_k) = \prod_{i \in s_{jk}} f_i \ .
\end{equation}
The indices in $s_{jk}$ correspond to the attributes based on which the encodings of the two symbols can be compared, and the values of these attributes are the components of the residue vectors $R(t_i)$ and $R(t_j)$ with indices in $s_{ik}$.

Let the reduced residue vector $R_{jk}(x)$ be obtained from the residue vector $R(x)$ with respect to $F$ by removing each component whose index does not appear in $s_{jk}$. Analogously,  let $F_{jk}$ be the relative prime vector obtained from $F$ by removing all components whose indices are not in $s_{jk}$. 
From the requirement of $s_j$ and $s_k$ to be lexicographically minimum index sets it follows that $R_{jk}(t_j)$ and $R_{jk}(t_k)$ respectively is the residue vector of $t_i$ and $t_j$ with respect to $F_{jk}$. We can therefore apply Lemma~\ref{lemma:residuevector} and Equation~(\ref{eq:gcd}) to show that 
$$
t_j \equiv t_k\ (\bmod \ \text{gcd}(p_j,p_k)) \ \textrm{if and only if} \ R_{jk}(t_j) = R_{jk}(t_k) \ ,
$$
which proves the theorem. 
\hfill{$\Box{}$}
}
\abstractonly{
\noindent
\textbf{Proof sketch of Theorem~\ref{thm:wstopc}}. Using the above method to compute a Partial Coding Problem instance with value ranges $(f_1,\ldots,f_d)$ and symbols $s_1,\ldots,s_n$ from a Windows Scheduling Problem instance with periods $p_1,\ldots,p_n$, it remains to show that there is a feasible solution to the scheduling problem if and only if there is a feasible partial coding.

We define a mapping between any schedule with job start times $t_1,\ldots,t_n$ and the corresponding partial coding. For $j=1,\ldots,n$ symbol $s_j$ is encoded by $c_j$ defined as 
$$
c_j(i) = r_i(t_j) \ ,
$$
where $r_i(t_j)$ is the $i$th component of the residue vector $R(t_j)$ with respect to $F$. Using   Lemma~\ref{lemma:residuevector}, one can show that this mapping is bijective. It therefore suffices to prove that in a schedule two jobs $j,k$ collide if and only if in the corresponding encoding the symbols $s_j,s_k$ are not distinguishable based on the attributes they have in common. 

Consider the vector $F_{jk}$ that is composed from the components of $F$ whose indices appear both in $s_j$ and $s_k$. The partial residue vector $R_{jk}(x)$ is analogously composed from components of $R(x)$; note that $R_{jk}(t_j)$ and $R_{jk}(t_k)$ contain exactly the set of attribute values of $c_j$ and $c_k$ by which the two symbols $s_j,s_k$ can be compared.
 Observing that the product of all components of $F_{jk}$ is $\text{gcd}(p_j,p_k)$, and that $R_{jk}(x)$ is the residue vector of $x$ with respect to $F_{jk}$, it follows from Lemma~\ref{lemma:residuevector} that $R_{jk}(t_j) = R_{jk}(t_k)$ if and only if $t_j \equiv t_k\ (\bmod\ \text{gcd}(p_j,p_k))$. Thus, the two symbols are incomparable iff the necessary and sufficient condition of Lemma~\ref{lemma:collision} for the jobs to collide is met.
\hfill{$\Box{}$}
}

Having reduced the Windows Scheduling Problem to Partial Coding, we now present the opposite direction of reduction. The first step is to reduce the Binary Partial Coding Problem to the Windows Scheduling Problem. Subsequently, we show how to reduce general Partial Coding to the binary case.

\begin{theorem}
\label{thm:bpctows}
The Binary Partial Coding Problem can be reduced in polynomial time to the Windows Scheduling Problem with one machine, uniform job lengths and exact periods. The reduction transforms instances of the former problem with $d$ attributes to instances of the latter problem with a maximum period of~$d^{O(d)}$.
\end{theorem}
\fullonly{
\textbf{Proof.} The approach of the reduction is to use a relative prime vector $F$ containing the first $2d$ prime numbers and construct Windows Scheduling instances where the jobs' periods can be represented as products over subsets of $F$. The well-known Prime Number Theorem~\cite{selberg49} states that the $m$th prime number is asymptotically $m \ln m$,  which already implies the claim that the maximum period of the resulting scheduling instance is upper bounded by $O([(2d) \ln (2d)]^{2d}) = d^{O(d)}$.

Instances of the Partial Coding Problem whose attribute value ranges correspond to pairwise distinct prime numbers can be straightforwardly transformed into Windows Scheduling instances by defining the period of the $j$th job as the product of the value ranges of all attributes contained by the $j$th symbol. The resulting scheduling instance is equivalent, because the set of attribute value ranges constitutes a relative prime vector $F$, and when using this $F$ to transform the scheduling problem back into an equivalent Partial Coding instance, like shown in the context of Theorem~\ref{thm:wstopc}, we arrive exactly back at the original instance.

We therefore show how to transform a given Binary Partial Coding Problem instance into an equivalent Partial Coding instance with attribute value ranges that are pairwise distinct primes. In what follows we assume an arbitrary but fixed set $A$ of $2d$ pairwise distinct attribute value ranges and show how to transform an instance of the Binary Partial Coding Problem with $d$ binary attributes into an equivalent instance of the Partial Coding Problem with $2d$ attributes whose value ranges correspond to exactly those in~$A$.

The transformation proceeds in $d$ iterations, where in each iteration one binary attribute, say $i$, is transformed into a pair of non-binary attributes. For transforming binary attribute $i$, we pick two arbitrary value ranges $a,b$ from $A$ that have not been picked in a previous iteration. Let w.l.o.g. $a<b$, and let $S(i)$ (resp. $\bar{S}(i)$) be the set of symbols in the current Partial Coding instance that contain (resp. do not contain) attribute $i$. We emphasize that the sets $S(i)$ and $\bar{S}(i)$ are defined with respect to the current problem instance, so, in case additional symbols have already been added to the instance in previous iterations, $S(i)$ and $\bar{S}(i)$ potentially also include those additional symbols.

The transformation in the current iteration proceeds by (1)~removing attribute~$i$, (2)~adding two new attributes $\alpha$ and $\beta$ with value ranges $a$ and $b$, respectively, (3)~adding $\alpha$ to all symbols in $\bar{S}(i)$ and adding $\beta$ to all symbols in $S(i)$, and finally (4)~introducing $(a-1) \cdot (b-2)$ new auxiliary symbols, each containing only $\alpha$ and $\beta$.

To see that this iteration results in an equivalent problem instance, observe that in any feasible encoding the $(a-1) \cdot (b-2)$ new auxiliary symbols need to be distinguished from both the symbols in $S(i)$ and the ones in $\bar{S}(i)$. Distinguishing from the symbols in $\bar{S}(i)$ is only possible by attribute $\alpha$; therefore the auxiliary symbols can assume at most $a-1$ different values of that attribute. Analogously, the new auxiliary symbols can assume at most $b-1$ different values of attribute $\beta$.

In order to be feasible, the encodings of the auxiliary symbols in terms of $\alpha$ and $\beta$ must be pairwise different. So if less than $a-1$ different values of $\alpha$ are assumed by the auxiliary attributes, at most $(a-2)\cdot (b-1)$ different encodings are possible, which is less than the number of new auxiliary symbols due to the assumption that $a < b$. We conclude that in any feasible encoding the auxiliary symbols assume \emph{exactly} $a-1$ different values in terms of attribute~$\alpha$. The remaining attribute value of $\alpha$ has to be assumed by the symbols in $\bar{S}(i)$, which implies that no pair of symbols from that set can be distinguished by attribute $\alpha$. 

From the necessity of the auxiliary symbols to assume exactly $a-1$ different values of $\alpha$ follows that, in order to be pairwise distinguishable, it suffices that they assume $b-2$ different values of attribute $\beta$. However, they cannot assume less than $b-2$ different values in terms of that attribute, again because otherwise the number of different value combinations would be outnumbered by the auxiliary symbols This implies that the symbols in $S(i)$ can assume up to two different values of attribute $\beta$, which implies that from the point of view of the attributes in $S(i)$ and $\bar{S}(i)$ the new attribute $\beta$ is a binary attribute replacing the previous binary attribute $i$.
\hfill{$\Box{}$}
}
\abstractonly{
\textbf{Proof sketch.} 
Instances of the Partial Coding Problem whose attribute value ranges correspond to distinct prime numbers can be straightforwardly transformed into Windows Scheduling instances by defining the period of the $j$th job as the product of the value ranges of all attributes contained by the $j$th symbol. The resulting scheduling instance is equivalent, because the set of attribute value ranges constitutes a relative prime vector $F$, and when using this $F$ to transform the scheduling problem back into an equivalent Partial Coding instance, like shown in the context of Theorem~\ref{thm:wstopc}, we arrive back at the original instance.

The approach of the reduction is to use a relative prime vector $F$ containing distinct prime numbers and construct Windows Scheduling instances where the jobs' periods can be represented as products over subsets of $F$. The well-known Prime Number Theorem~\cite{selberg49} states that the $m$th prime number is asymptotically $m \ln m$,  which already implies the claim that the maximum period of the resulting scheduling instance is upper bounded by $O((d \ln d)^d) = d^{O(d)}$.

To transform a Binary Partial Coding Problem instance into an equivalent instance with attribute value ranges that are pairwise distinct primes, we use a technique to replace a binary attribute~$i$ with a pair of non-binary attributes having arbitrary value ranges. Let $a$  and $b$ be those value ranges, and let w.l.o.g. $a<b$. Let $S(i)$ (resp. $\bar{S}(i)$) be the set of symbols in the current Partial Coding instance that contain (resp. do not contain) attribute~$i$. The transformation proceeds by (1)~removing attribute~$i$, (2)~adding two new attributes $\alpha$ and $\beta$ with value ranges $a$ and $b$, respectively, (3)~adding $\alpha$ to all symbols in $\bar{S}(i)$ and adding $\beta$ to all symbols in $S(i)$, and finally (4)~introducing $(a-1) \cdot (b-2)$ new auxiliary symbols, each containing only $\alpha$ and $\beta$. By reasoning about the necessity to distinguish the new auxiliary symbols from each other and from all other symbols, one can show that in any feasible encoding the auxiliary symbols must assume exactly $a-1$ different values of attribute $a$ and at least $b-2$ different $\beta$-values. The remaining two values of $\beta$ can be assumed by the symbols in $S(i)$ and therefore $\beta$ replaces $i$ as a binary attribute. 
\hfill{$\Box{}$}
}

The computational equivalence of the Windows Scheduling Problem and the Partial Coding Problem is completed by giving a reduction of the general Partial Coding Problem to the binary case.

\begin{lemma}
\label{lemma:pctobpc}
The Partial Coding Problem can be reduced to the Binary Partial Coding Problem in polynomial time.
\end{lemma}
\fullonly{
\textbf{Proof.} Let $F = (f_1,\ldots,f_d)$ be the attribute value ranges and let $s_1,\ldots,s_n$ be the symbols of a Partial Coding Problem instance. We show how to replace a single non-binary attribute $f_i$ with a set of binary attributes. This procedure can then be applied to all non-binary attributes, arriving at an instance of the Binary Partial Coding Problem.

For some $f_i > 2$, let $k := \lceil \log f_i \rceil$. Let $S(i)$ be the set of symbols containing attribute~$i$, and let $\bar{S}(i)$ be the set of symbols not containing it.
We remove attribute~$i$ and introduce instead $k + 1$ new binary attributes. Out of these binary attributes, all but one are used as \emph{binary encoding attributes} to express the value of the former attribute~$i$ in binary form. Accordingly, index $i$ is replaced in each symbol $s_j \in S(i)$ with the indices of those $k$ binary attributes. Note that the $k$ attributes can assume $2^k$ different value combinations, which is less than twice the value range of attribute~$i$.

To reduce the number of combinations back to $f_i$, we introduce a number of auxiliary symbols. Let $m := 2^k - f_i$, and consider the representation of $m$ as a binary number, that is, $m = \sum_{j=0}^{k-1} c_j 2^j$ with $c_j \in \{0,1\}$ for $j = 0,\ldots,k-1$. For each $j$ with $c_j = 1$ we introduce an auxiliary symbol that contains the first $k-j$ auxiliary attributes (assuming an arbitrary but fixed order of these attributes). Figure~\ref{fig:pctobpc} shows an example of the transformation.

Recall that only $k$ of the $k+1$ new attributes have been added to symbols yet. We add the  remaining attribute to all new auxiliary symbols and to all symbols in $\bar{S}(i)$. By this single binary attribute, the new auxiliary symbols can be distinguished from the symbols in $\bar{S}(i)$. In any feasible encoding, all new auxiliary symbols must assume one value of that binary attribute, and the symbols in $\bar{S}(i)$ must assume the other, so that attribute is neither available to distinguish auxiliary symbols from each other, nor for distinguishing symbols in $S(i) \cup \bar{S}(i)$ from each other.

For $j = 0,\ldots,k-1$, if $c_j = 1$, then the corresponding auxiliary symbol must be distinguished from all other symbols except those in $\bar{S}(i)$ based on the first $k-j$ binary encoding attributes. Given any encoding, let $(v_1,\ldots,v_{k-j})$ be the binary vector of those attribute values of the auxiliary symbol. Define vector set $V_j$ as the set of all length $k$ binary vectors having $(v_1,\ldots,v_{k-j})$ as a prefix. Clearly, $|V_j| = 2^j$. As the auxiliary symbols must be pairwise distinguishable, it follows that $V_j \cap V_{j^\prime} = \emptyset$ for $j \neq j^\prime$. Therefore $\sum_{c_j = 1} |V_j| = m$. As all symbols in $S(i)$ must be distinguished from the auxiliary symbols in terms of the $k$ binary encoding attributes, no symbol in $S(i)$ can be assigned an attribute value combination appearing in any $V_j$, so the number of different attribute value combinations available for the symbols in $S(i)$ is $2^k - m = f_i$, which shows equivalence to the original problem instance.

To analyze the runtime of this method, observe that each time a non-binary attribute~$i$ is eliminated like shown above, both the number of attributes and the number of symbols increases by no more than $\lceil \log f_i \rceil$.  After having eliminated all non-binary attributes, the resulting instance of the Binary Partial Coding Problem therefore has at most $O (d \log f_{\max})$ attributes and $n + O(d \log f_{\max})$ symbols, where $f_{\max}$ is the maximum attribute value range in the original instance. 
\hfill{$\Box{}$}
}
\abstractonly{
\textbf{Proof sketch.} The idea of the proof is to replace each attribute having value range $f_i$ with $O(\log f_i)$ binary attributes, so that the number of possible value combinations the binary attributes can assume is $f_i$. When this is not directly possible because $f_i$ is no power of $2$, a set of auxiliary symbols is added to limit the number of value combinations that remain available for the non-auxiliary symbols to $f_i$. We refer the reader to the full article for details on the construction; see also Figure~\ref{fig:pctobpc} for an example.
\hfill{$\Box{}$}
}
\begin{figure}[h]
\begin{minipage}{0.5 \textwidth}
\center
\begin{tabular}{l|cc}
{} & $s_1$ & $s_2$
\\
\hline
\{0,1,2\} & $\times$ & $\times$ 
\\
\{0,\ldots,10\} & $$ & $\times$ 
\end{tabular}
\end{minipage}
\begin{minipage}{0.5 \textwidth}
\center
\begin{tabular}{l|cccc}
{} & $s_1$ & $s_2$ & $a_1$ & $a_2$
\\
\hline
\{0,1,2\} & $\times$ & $\times$ 
\\
\{0,1\} & $$ & $\times$ & $\times$ & $\times$
\\
\{0,1\} & $$ & $\times$ & $\times$ & $\times$
\\
\{0,1\} & $$ & $\times$ & $$ & $\times$
\\
\{0,1\} & $$ & $\times$ & $$ & $\times$
\\
\{0,1\} & $\times$ & $$ & $\times$ & $\times$
\end{tabular}
\end{minipage}
\caption{
\label{fig:pctobpc}
The non-binary attribute with value range $11$ in the left-hand problem instance is replaced with a set of $\lceil \log_2 11 \rceil + 1 = 5$ binary attributes and two auxiliary symbols in the right-hand instance. The top $4$ binary attributes are used for binary encoding of the previous non-binary attribute values, while the two auxiliary symbols reduce the number of available value combinations from 16 back to~11. The bottom binary attribute is used for making the auxiliary symbols comparable with symbol $s_1$.
}
\end{figure}

\begin{corollary}
\label{cor:pctows}
The Partial Coding Problem can in polynomial time be reduced to the Windows Scheduling Problem with one machine, uniform job lengths and exact periods.
\end{corollary}

\section{Single-machine Windows Scheduling}
\label{sec:hardness}

In this section we address the computational complexity of Windows Scheduling with unit-length jobs and a single machine. 
We first show that the Binary Partial Coding Problem with exact periods is NP-hard. This is done by a fairly simple proof of NP-completeness of the Partial Coding (PC) Problem by reduction from Graph Coloring. As the reduction from BPC to WS proved as Theorem~\ref{thm:bpctows} results in WS instances having exponentially large periods, this simple reduction shows only weak NP-hardness of Windows Scheduling, which is not a new result because a proof of weak NP-hardness has already been given in~\cite{bar02} by a direct reduction from Graph Coloring. 
The reason why we provide the reduction to PC is because it serves as the starting point of the refined reduction presented thereafter.

\begin{definition}
Give a graph $G=(V,E)$ and a number $k$ of colors, the Graph Coloring Problem is to decide whether each node in $V$ can be assigned one of the $k$ colors such that for any edge $(v,w) \in E$ the color of $v$ is different to the color of $w$.
\end{definition}

The Graph Coloring Problem is NP-complete~\cite{garey79}, so reducing it to the Binary Partial Coding problem shows NP-hardness of the Windows Scheduling Problem. 

\begin{lemma}
\label{lemma:gctopc}
The Partial Coding Problem is NP-Complete.
\end{lemma}
\fullonly{
\textbf{Proof.} The feasibility of any given encoding can be straightforwardly verified, showing that Partial Coding is in NP. To show NP-hardness, we reduced instances $G=(V,E)$ of the $k$-coloring problem to Partial Coding instances as follows. Let $V = \{v_1,\ldots,v_n\}$. For $j=1,\ldots,n$ we introduce one symbol $s_j$. For every pair of nodes $v_i,v_j \in V$ that are \emph{not} connected by an edge, we define a binary attribute $a_{ij}$ and add it only to the two symbols $s_i,s_j$. In addition, we introduce one attribute $a^*$ with value range $k$ and add it to all $n$ symbols.

A feasible $k$-coloring can be transformed into an encoding by assigning to each symbol $s_i$ the $a^*$-value corresponding to the color of node $v_i$. Furthermore, each binary attribute $a_{ij}$ is used to distinguish the two symbols $s_i$ and $s_j$ by assigning them distinct values of that attribute. Consequently, every symbol pair $s_i,s_j$ with $(v_i,v_j) \notin E$ is distinguished by attribute $a_{ij}$ and every symbol pair $s_i,s_j$ with $(v_i,v_j) \in E$ is distinguished by attribute $a^*$ due to the feasibility of the coloring. 

If there is no feasible $k$-coloring, then there is no encoding which distinguishes all symbol pairs $s_i,s_j$ with $(v_i,v_j) \in E$ based on attribute $a^*$. As $a^*$ is the only attribute these symbol pairs have in common, it follows that there is no feasible encoding.
\hfill{$\Box{}$}
}
\abstractonly{
\textbf{Proof sketch.} To show NP-hardness, we reduce from instances $G=(V,E)$ of the $k$-coloring problem. Let $V = \{v_1,\ldots,v_n\}$. For $j=1,\ldots,n$ we introduce one symbol $s_j$. For every pair of nodes $v_i,v_j \in V$ that are \emph{not} connected by an edge, we define a binary attribute $a_{ij}$ and add it only to the two symbols $s_i,s_j$. In addition, we introduce one attribute $a^*$ with value range $k$ and add it to all $n$ symbols. The equivalence of the Graph Coloring instance and the Partial Coding instance is not hard to verify; details can be found in the full article.
\hfill{$\Box{}$}
}

\begin{corollary}
The Binary Partial Coding Problem and the one-machine Windows Scheduling Problem with unit length jobs and exact periods are NP-complete.
\end{corollary}
\fullonly{
\textbf{Proof.} Membership of BPC in NP is straightforward, and NP-hardness follows from Lemma~\ref{lemma:gctopc} just proved and the reduction given in Lemma~\ref{lemma:pctobpc}. Membership of Windows Scheduling in NP follows from the fact that the collision of jobs can be checked using Lemma~\ref{lemma:collision}, while NP-hardness follows from the reduction stated in Theorem~\ref{thm:bpctows}. 
\hfill{$\Box{}$}
}

In what follows we tighten this hardness result by showing that also pseudo-polynomial algorithms for one-machine Windows Scheduling with unit-length jobs are unlikely to exist. 

\begin{theorem}
\label{thm:constgctobpc}
Instances of the $3$-Graph Coloring Problem with $n$ nodes having constant degree can be reduced to instances of the Binary Partial Coding problem consisting of $O(n)$ symbols and $O(\log^2 n)$ attributes in expected polynomial time.
\end{theorem}
\fullonly{This theorem implies the improved complexity result stated as follows.}
\abstractonly{In combination with Theorem~\ref{thm:bpctows} and the fact that Graph Coloring is NP-hard even for constant degree graphs~\cite{dailey80}, this implies the following complexity result.}

\begin{corollary}
\label{cor:gctows}
The Windows Scheduling Problem does not admit a pseudo-polynomial time algorithm unless SAT can be solved by a randomized algorithm in expected time $n^{O(\log n \log \log n)}$.
\end{corollary}
\fullonly
{
\textbf{Proof.} The combination of Theorem~\ref{thm:constgctobpc} and Theorem~\ref{thm:bpctows} shows that instances of $3$-Graph Coloring with constant degree graphs having $n$ nodes can be reduced to instances of the Windows Scheduling Problem, having a maximum period length of $(\log^2 n)^{O(\log^2 n)} = n^{O(\log n \log \log n)}$, by a randomized method in expected polynomial time. A pseudo-polynomial time algorithm for the Windows Scheduling Problem could solve that instance in time $n^{O(\log n\log\log n)}$.
As Graph Coloring is NP-hard for constant degree graphs~\cite{dailey80}, SAT can be reduced to it in polynomial time, implying the claim.
\hfill{$\Box{}$}
}

\fullonly{
\noindent
\textbf{Proof of Theorem~\ref{thm:constgctobpc}.}
In the proof of Lemma~\ref{lemma:gctopc}, Graph Coloring has been reduced to Partial Coding by defining for each unconnected node pair a binary attribute by which the corresponding two symbols in the Partial Coding instance can be compared. The resulting number of attributes is quadratic in the number of graph nodes. The idea of this proof is to use fewer attributes to represent the set of non-edges (i.e. node pairs not connected by an edge). 

Consider a set of $m$ symbols, and assume that these symbols share an attribute having value range $m$. The $m$ symbols can be all distinguished from each other by this attribute. If these $m$ symbols correspond to $m$ nodes that form an independent set in the Graph Coloring instance, this single attribute is an efficient way to cover about $m^2$ non-edges. Using the idea of more efficient coverings, we try to cover the set of all non-edges in the graph by a small number of independent sets. In other words, our reduction finds a small \emph{edge clique cover} for the complement graph of the Graph Coloring instance.

While edge coverings with few cliques might not exist for general graphs, we make use of a result given in~\cite{alon86}. In that work it is shown that edge clique covers of size $O(\log n)$ exist in complements of constant degree graphs. The proof uses a probabilistic argumentation and is not directly constructive, but we can use the same approach in a randomized construction.

Let $G=(V,E)$ be the given Graph Coloring instance with $|V|=n$ and constant maximum degree $c$. For the sake of simplicity we assume that the number admitted of colors is fixed to $4$ instead of $3$, which is admissible as the case of $3$ colors be straightforwardly be reduced to $4$-Graph Coloring.

For some value of $m$ to be determined later, the algorithm generates $m$ independent sets $S_1,\ldots,S_m$ by a randomized method. Each of these sets $S_i$ is generated by first choosing each node in $V$ with a fixed probability $p$ and then removing from $S_i$ each $v$ with $(v,v^\prime) \in E$ for some $v^\prime \in S_i$. 
The resulting set $S_i$ is an independent set in $G$.
Each node $v$ is chosen into $S_i$ with probability $p$ and remains in $S_i$ with probability at least $(1-p)^c$. Therefore, each non-edge $(v,v^\prime)$ is covered by the independent set $S_i$ with a constant probability $p_e$ of at least $p^2(1-p)^{2c}$. 

The probability that a given non-edge is not contained in any of the $m$ independent sets is at most $(1 -p_e)^m$. We choose $m$ such that 
\begin{equation}
\label{eq:m}
(1-p_e)^m < \frac{1}{n^2} , 
\end{equation}
which, due to $p_e$ being a constant independent of $n$, can be achieved by some $m = O(\log n)$. As a result of Equation~(\ref{eq:m}), the expected number of non-edges not covered by  $S_1,\ldots,S_m$ less than $1$. This is as a proof for the existence of a size $m$ cover of all non-edges. Furthermore, it follows from this expected value that the probability that all edges are covered by the $m$ sets is larger than $\frac{1}{2}$. 

The procedure of generating $m$ sets is repeated until a cover of size $m$ has been found. As the success probability is larger than $\frac{1}{2}$, the expected number of tries is less than $2$.

After that, an instance of the Partial Coding Problem is constructed having $n$ symbols that correspond to the $n$ nodes of $G$. For each $i = 1,\ldots,m$ an attribute $a_i$ with value range $n^\prime$ is added, where $n^\prime = \min\{2^k | 2^k > n\}$. This attribute is added to each symbol where the corresponding graph node appears in $S_i$. Furthermore, we add an additional attribute having value range $4$ to all $n$ symbols. 

The constructed Partial Coding instance is equivalent to the Graph Coloring instance, because every pair of symbols corresponding to a non-edge can be distinguished based on some of the attributes $a_1,\ldots,a_m$, while every pair of symbols corresponding to an edge must be distinguished by the value range $4$ attribute. 

To transform this Partial Coding instance into a Binary Partial Coding instance, we replace each attribute with value range $n^\prime$ with $\log n^\prime$ binary attributes, and we replace the value range $4$ attribute with $2$ binary attributes. Here the fact that the Partial Coding instance has only attributes whose value range are powers of $2$ becomes beneficial, because there is no need for auxiliary symbols like in previously given reductions (see proof of Lemma~\ref{lemma:pctobpc}). The resulting Binary Partial Coding instance has $n$ symbols and $m \log n^\prime + 2 = O(\log^2 n)$ attributes, which concludes the proof.
\hfill{$\Box{}$}
}
\abstractonly{
\noindent
\textbf{Proof sketch of Theorem~\ref{thm:constgctobpc}.}
In the proof of Lemma~\ref{lemma:gctopc}, Graph Coloring has been reduced to Partial Coding by defining for each unconnected node pair a binary attribute by which the corresponding two symbols in the Partial Coding instance can be compared. The resulting number of attributes is quadratic in the number of graph nodes. The idea of this proof is to use fewer attributes to represent the set of non-edges (i.e. node pairs not connected by an edge). 

Consider a set of $m$ symbols, and assume that these symbols share an attribute having value range $m$. The $m$ symbols can be all distinguished from each other by this attribute. If these $m$ symbols correspond to $m$ nodes that form an independent set in the Graph Coloring instance, this single attribute is an efficient way to cover about $m^2$ non-edges. Using the idea of more efficient coverings, we try to cover the set of all non-edges in the graph by a small number of independent sets. In other words, our reduction finds a small \emph{edge clique cover} for the complement graph of the Graph Coloring instance.

While edge coverings with few cliques might not exist for general graphs, we make use of a result given in~\cite{alon86}. In that work it is shown that edge clique covers of size $O(\log n)$ exist in complements of constant degree graphs. The proof uses a probabilistic argumentation and is not directly constructive, but one can use the same approach in a randomized construction to find an edge clique cover of size $O(\log n)$ in expected polynomial time; details can be found in the full article.
\hfill{$\Box{}$}
}

We now consider the one-machine Window Scheduling Problem with $n$ unit-length jobs and \emph{inexact} periods. Recall that here each job $j$ has to be scheduled such that the time interval between two consecutive executions is no longer that $p_j$. A well-known necessary condition for feasibility is that the \emph{density} of the instance must be less than one, where (for the special case of unit-length jobs) the density is defined as $\sum_{j=1}^{n} \frac{1}{p_j}$. The intuition behind this condition is that a job with period $p_j$ requires a share of $1/p_j$ of all available time units, so the total share of all jobs must be less or equal to $1$. Furthermore, problem instances having a density of \emph{exactly} $1$ must be scheduled such that the time interval between any two consecutive executions of job $j$ is exactly $p_j$. Here the intuitive explanation is that only with exact periods a job $j$ does not occupy a larger share than $\frac{1}{p_j}$ of all time units. Formal proofs can be found in~\cite{holte89}.

Using this property of density $1$ instances, hardness proofs for the problem with inexact periods can be obtained by showing NP-hardness of density $1$ instances, as in this class of instances the periods can as well be assumed to be exact. The approach that has been taken in~\cite{bar07} is to augment  instances of the problem with exact periods by additional jobs such that the density becomes $1$. The period of all these additional jobs is chosen as the least common multiple $v$ of all original jobs' periods. As any overall schedule is periodic with period $v$ (see~\cite{holte89}), the additional jobs can be inserted anywhere in a feasible schedule for the original instance, and so the resulting problem instance admits a feasible schedule if and only if the original instance is solvable.

The number of additional jobs used in that approach is exponential. As they are all identical, the resulting instance can still be represented in a compact way using polynomial space. The compact  representation specifies for each period the number of jobs having that period. Therefore, the proof of weak NP-hardness given in~\cite{bar02} for exact periods implies weak NP-hardness for the compact representation of the inexact periods case.

\fullonly{
In what follows we use the complexity analysis given earlier in this section to make a statement on the computational complexity of the Windows Scheduling problem with inexact periods in standard representation.
}
\abstractonly{
Using the complexity analysis given earlier in this section, we can reduce the number of additional jobs to $n^{O(\log n \log \log n)}$. The formal proof of the resulting theorem is omitted in this extended abstract.
}

\begin{theorem}
The single-machine Windows Scheduling Problem with unit-length jobs and inexact periods does not admit a pseudo-polynomial time algorithm unless SAT can be solved in expected time $n^{O(\log n \log \log n)}$.
\end{theorem}
\fullonly{
\textbf{Proof.} The reasoning in the proof of Corollary~\ref{cor:gctows} was to show that a $3$-Graph Coloring instance with constant degree and $n$ nodes can be reduced to to an instance of the Windows Scheduling problem having a maximum period of $n^{O(\log n \log \log n)}$. The periods have been obtained as products over a subset of members of a relative prime vector $F$, and the upper bound on the period has been calculated as the product of all members of $F$. It follows that the least common multiple of all periods is $q = n^{O(\log n \log \log n)}$.

Every schedule for the obtained scheduling instance is periodic with period of $q$, i.e., the schedule repeats every $q$ time steps. For each job $j$ there are $\frac{q}{p_j}$ time slots within the interval $[0,q-1]$ where it is executed. Therefore, in any feasible schedule, the number of idle time slots within $[0,q-1]$ is $r := q - \sum_j \frac{q}{p_j}$. So when we modify the problem instance by adding $r$ additional jobs with period $q$, we arrive at a problem instance which admits a feasible schedule if and only if the original problem instance does. As the new problem instance has a density of $1$, the existence of a feasible schedule for it does not depend on whether or not the periods are exact. 

As $r$ is upper bounded by $q = n^{O(\log n \log \log n)}$, the resulting Windows Scheduling instance with inexact periods has $n^{O(\log n \log \log n)}$ jobs and a maximum period of $n^{O(\log n \log \log n)}$. A pseudo-polynomial algorithm for the Windows Scheduling Problem with inexact periods could therefore solve it in time $n^{O(\log n \log \log n)}$, and the solution would decide the NP-hard Graph Coloring problem we have started the reduction from.
\hfill{$\Box{}$}
}

\section{Multi-machine Windows Scheduling}
\label{sec:multimachine}

We now address the multi-machine case of the Windows Scheduling problem. Being a generalization of the single-machine case, all hardness results shown so far extend to it. As pointed out in Section~1, there are two variants of the problem with multiple machines. In the version where \emph{job migration} is disallowed, there has to be a fixed assignment of jobs to machines, and each machine has to individually process the jobs that have been assigned to it. This problem version is not harder than the single machine case, as shown later.

We start by a remark on the proof of NP-hardness given in~\cite{bar07} for the case of unit length jobs and $h \geq 1$ machines without machine migration, which we believe is incorrect. The authors reduce from the 3-Dimensional Matching Problem (3DM), where for a set of triplets $T \subset X \times Y \times Z$ with $|X| = |Y| = |Z| = h $ the problem is to decide whether there is a subset of $T$ consisting of exactly $h$ non-overlapping triplets. In the given reduction, each triplet in $T$ is assigned a distinct prime number larger than 2, and each element of $X \cup Y \cup Z$ is interpreted as a job whose period is obtained as the product the prime numbers of all triplets the element appears in. The correctness of the reduction is based on the claim that three jobs can be processed by the same machine if and only if they appear in the same triplet, making use of the collision condition that appears in this work as Lemma~\ref{lemma:collision}.

However, the \emph{only if} implication does not hold true, which is demonstrated by a simple example. Consider $h = 2$ and let $X = \{x_1,x_2\}, Y = \{y_1,y_2\}, Z=\{z_1,z_2\}$ and $T = \{(x_1,y_1,z_1),(x_1,y_2,z_2),(x_2,y_1,z_2),(x_2,y_2,z_1)\}$. This 3DM instance has no solution because every triplet pair overlaps at some variable. However, $\{x_2,y_2,z_2\}$ can be scheduled on the same machine: Assuming that the triplets in $T$ are assigned the prime numbers $3,5,7,11$, it follows that $\text{gcd}(x_2,y_2) = 11, \text{gcd}(x_2,z_2) = 7, \text{gcd}(y_2,z_2) = 5$. If we assign a start time of $1,2,3$ respectively to $x_2,y_2,z_2$, Lemma~\ref{lemma:collision} can be applied to each pair among these three jobs, verifying that there is no collision.

Despite this error, the problem remains NP-hard as it is a generalization of the single-machine problem addressed in the previous section. In what follows we prove that multi-machine Windows Scheduling without machine migration is not harder than its special case with a single machine.

\begin{theorem}
\label{thm:multitosingle}
The multi-machine Windows Scheduling Problem with unit-size jobs having exact periods and no machine migration allowed can be reduced to the single-machine variant of that problem in polynomial time.
\end{theorem}
\fullonly{
\textbf{Proof.} We transform the Windows Scheduling instance with $m$ machines to a Partial Coding problem instance as shown as the proof of Theorem~\ref{thm:wstopc}. The Windows Scheduling instance admits a schedule for $m$ machines if and only if the set of symbols in the Partial Coding instance can be partitioned into $m$ subsets such that for each of these symbol subsets a feasible encoding exists. This is equivalent to introducing a new attribute with value range $m$, adding it to all symbols, and considering the resulting problem as a standard instance of Partial Coding. The reduction is then completed by transforming the resulting Partial Coding instance back to Windows Scheduling as stated in Corollary~\ref{cor:pctows}.
\hfill{$\Box{}$}
}
\abstractonly{
\abstractonly{
\textbf{Proof sketch.}  Windows Scheduling with $m$ machines without migration can be expressed as a Partial Coding problem where the set of symbols needs to be divided into $m$ subsets and then a feasible encoding for each subset is to be found. This is equivalent to the introduction of an additional attribute having value range $m$.
\hfill{$\Box{}$}
}
}

In the second variant of the multi-machine case, machine migration is permitted. 
Here a schedule is feasible if in any time slot there are at most $m$ jobs executed. 
Unlike the problem version without machine migration, this case version turns out to be computationally harder than the single-machine variant. 

Similarly to a hardness proof given in~\cite{baruah90} for a related periodic scheduling problem, we apply the Generalized Chinese Remainder Theorem to arrive at the following characterization of infeasible schedules. 

\begin{lemma}
\label{lemma:crt}
Given an instance of the Windows Scheduling Problem with $m$ machines, $n$ unit-size jobs with exact periods $p_1,\ldots,p_n$ and machine migration allowed, a given schedule is infeasible if and only if there exists a set $K$ of $m+1$ jobs such that each pair of jobs from $K$ collides in that schedule.
\end{lemma}
\abstractonly{A proof can be found in the full version of this article.}
\fullonly{
\textbf{Proof.}  If $t_1,\ldots,t_n$ are start times of the $n$ jobs, then the schedule is infeasible iff there exists some subset $K$ of $m+1$ jobs and some time $t^*$ such that $t^* \equiv t_j \ (\bmod p_j) $ for $j \in K$. From the Generalized Chinese Remainder Theorem~\cite{knuth14} follows that this is the case if and only if for each pair $i,j \in K$ it holds that $t_i \equiv t_j\ (\bmod \text{gcd}(p_i,p_j)$. The latter property is stated in Lemma~\ref{lemma:collision} as necessary and sufficient for jobs $i$ and $j$ to collide in the schedule.
\hfill{$\Box{}$}
}

\begin{theorem}
\label{thm:conp}
The Windows Scheduling Problem with $m$ machines, exact periods, and machine migration is 
co-NP-hard even for unit-length jobs.
\end{theorem}
For proving the theorem, we introduce a generalization of the Partial Coding Problem which corresponds to the multi-machine Windows Scheduling Problem with job migration. 

\begin{definition}
Given attributes with value ranges $(f_1,\ldots,f_d)$ and symbols $s_1,\ldots,s_n \subseteq \{1,\ldots,d\}$, the $k$-ary Partial Coding Problem is to determine whether there is an encoding of the symbols in terms of their attributes such that any subset of more than~$k$ symbols contains two elements that are distinguishable by the values of their common attributes.
\end{definition} 

\fullonly{
\noindent
\textbf{Proof of Theorem~\ref{thm:conp}.} In context of the proof of Theorem~\ref{thm:wstopc} we have shown that an instance of the one-machine Windows Scheduling Problem can be transformed into a Partial Coding instance based on a relative prime factorization $F$ of the multi-set of job periods. A bijective mapping between encodings of symbols and start times of jobs has been defined, with the property that two jobs collide if and only if the corresponding two symbols in the Partial Coding instance are not distinguishable. 

The same reduction can be used to transform an instance of the $m$-machine Windows Scheduling Problem into an instance of the $m$-ary Partial Coding Problem. The problem instances are equivalent, as Lemma~\ref{lemma:crt} establishes equivalence of the common collision of a set of $m+1$ jobs with the pairwise collision of any two members of that set. 

In the proof of Theorem~\ref{thm:bpctows} we have already used the fact that every Partial Coding instance whose attribute value ranges are pairwise distinct prime numbers can be directly reduced back into a Windows Scheduling instance. Consequently, these kind of instances of the $m$-ary Partial Coding Problem have equivalent $m$-machine Windows Scheduling instances. It therefore suffices to show co-NP-hardness of $m$-ary Partial Coding with attribute value ranges that are distinct primes.

We reduce from the $(m+1)$-Independent Set Problem, where for a given graph $G=(V,E)$ the task is to decide whether $V$ contains a subset of size $m+1$ whose members are pairwise not connected. The Partial Coding instance constructed by the reduction contains a symbol $s_i$ for each node $v_i \in V$ and an attribute for each edge. The attribute corresponding to an edge $(v_i,v_j)$ is only contained by the symbols $s_i,s_j$. The value ranges of the attributes are chosen as the first $|V|$ prime numbers. 

As each attribute can assume at least $2$ values and has only two symbols containing it, these two symbols can be distinguished by that attribute, and doing so has no influence on whether other pairs of symbols can be distinguished. This means that whether two symbols can be distinguished only depends on whether there is an edge between the corresponding pair of vertices.
Therefore, an independent set of size $m+1$ exists in the graph if and only if there is a set of $m+1$ symbols that pairwise cannot be distinguished in any solution of this Partial Coding instance. As the Independent Set Problem is NP-hard, it follows that $m$-ary Partial Coding is co-NP hard for attribute value ranges that are pairwise distinct primes, implying co-NP-hardness of $m$-machine Windows Scheduling with machine migration.
\hfill{$\Box{}$}
}
\abstractonly{
\noindent
\textbf{Proof sketch of Theorem~\ref{thm:conp}.} The idea of the proof is to show co-NP-hardness of the $k$-ary Partial Coding Problem by reduction from the $(k+1)$-Independent Set Problem, where for a given graph $G=(V,E)$ the task is to decide whether $V$ contains a subset of size $k+1$ whose members are pairwise not connected. The Partial Coding instance constructed by the reduction contains a symbol $s_i$ for each node $v_i \in V$ and an attribute for each edge. The attribute corresponding to an edge $(v_i,v_j)$ is only contained by the symbols $s_i,s_j$. It can be shown that an independent set of size $k+1$ exists if and only if there is a set of $k+1$ symbols that pairwise cannot be distinguished in any solution of this Partial Coding instance. 
\hfill{$\Box{}$}
}

\section{Summary and Conclusion}

In this work we have identified the Partial Coding Problem as the combinatorial core of the Windows Scheduling Problem. As polynomial-time reductions in both directions exist, the new interpretation is applicable for deriving both new algorithmic results and hardness proofs for Windows Scheduling. The focus of this work has been on the latter possibility. We have proved that even the single-machine case with uni-length jobs cannot be solved in pseudo-polynomial time under standard complexity-theoretic assumptions, and we have shown co-NP-hardness of the multi-machine case with machine migration. Remarkably, the latter problem is both NP-hard and co-NP-hard, which implies that it belongs to neither NP nor co-NP unless NP=co-NP.

We believe that the Partial Coding interpretation has the potential to yield further positive results in addition to the reduction of a multi-machine version to the single machine case given in this work. A problem that is currently open is the existence of an efficient algorithm for the class of problem instances with constantly many different jobs, possibly generalizing the algorithms for up to $3$ different jobs given in~\cite{holte89,holte92}. We furthermore see a potential application of our new interpretation in the study of approximation algorithms for Windows Scheduling.

\bibliography{relatedwork}

\end{document}